# Importance of Non-Perturbative QCD Parameters for Bottom Mesons

A. Upadhyay, M. Batra
School of Physics and Material Science
Thapar University, Patiala, Punjab-147004, India

**Abstract**
The importance of non-perturbative Quantum Chromodynamics [QCD] parameters is discussed in context to the predicting power for bottom meson masses and isospin splitting. In the framework of heavy quark effective theory, the work presented here focuses on the different allowed values of the two non-perturbative QCD parameters used in heavy quark effective theory formula and using the best fitted parameter, masses of the excited bottom meson states in $j^p = \frac{1}{2}^+$ doublet in strange as well as non-strange sector are calculated here. The calculated masses are found to be matching well with experiments and other phenomenological models. The mass and hyperfine splitting has also been analyzed for both strange and non-strange heavy mesons with respect to spin and flavor symmetries.

## 1. Introduction

An interesting class of non-perturbative Quantum Chromodynamics, now a day, is the study of mesons containing a heavy quark with a light meson using heavy quark effective theory [HQET]. In the light of heavy quark effective theory, spin and parity of the heavy quark decouples from that of light quark. Thus, the properties of heavy hadrons are independent of spin and flavor of heavy meson and HQET provides a basis for estimation of several properties of heavy mesons. HQET provides a systematic expansion of mass of heavy quark in terms of QCD parameters $\bar{\Lambda}/m_Q$. In the present work, masses of bottom mesons for excited states are predicted using $O(1/m_Q)$ HQET formula which contains terms with mass of heavy quark and two matrix elements of HQET operators. The motivation for present work arises due to new revolution in Charmonia with BaBar's discovery of a narrow meson, $D_{0s}^+$ (2317) [1]. Soon, after the discovery of $D_{0s}^+$ (2317) state, Focus [2] and CLEO [3] confirmed the same resonance along with observation of another narrow state $D_{1s}^+$(2463). Both these states were confirmed later by Belle [4]. More recently other candidates have been added to the list $D_{sJ}$(2860) by BaBar [5] and $D_{sJ}$(2710) by BaBar and Belle both [5, 6]. Since then CLEO [3], Belle [6], Fermilab [7] & BES have predicted many new states which created great enthusiasm in the Charm sector. The status of bottom meson spectroscopy on experimental grounds is slightly at a lower position than charm mesons. Some p-wave bottom and bottom strange mesons have been discovered earlier by collaborations like DELPHI [9] and ALEPH [10]. D0 has also observed evidence for the $B^*_{s2}$ meson at a mass of (5839.1MeV) [11]. This value has also been confirmed by CDF with higher precision [10]. But there are other measurements of excited B mesons masses $B_{s1}$ (5829.4±0.7) and $B_{s2}^*$ [11, 12] reported by CDF and D0 which differ significantly and more data are needed to get precise masses and widths. V. M. Abazov (D0 Collaboration) [13] in 2008 presented first strong evidence for resolution of excited B mesons $B_1$ and $B_2^*$. The mass of $B_1$ is measured to be 5720±1.4 MeV/c$^2$ and $B_2^*$=5746.8±2.4±1.7 MeV/c$^2$. Very recently, the new states have been predicted for orbitally excited bottom mesons by CDF collaboration. The mass of this new resonance has been found to be $5978\pm5\pm12\, MeV/c^2$ for neutral states and $5961\pm5\pm12\, MeV/c^2$ for charged state [14]. Thus, a lot of experiments have started to look at charm and bottom mesons spectrum and hence the old theories have also been revived. One of these approaches is heavy quark effective theory 'HQET' [15-17] to study heavy-light hadrons. For heavy quarks, it is possible to parameterize in HQET, the non-perturbative affects to a given order in $1/m_Q$ expansion in terms of a few unknown constants where these unknown constants can be obtained from the experiments. These un-known constants are actually the non-perturbative QCD parameters and once their values in hand, it is possible to calculate masses of various excited states in the heavy meson spectrum. We find it more suitable to obtain bottom meson spectrum due to lesser values of mass splitting for bottom mesons than that of charm mesons. B meson masses in the heavy quark effective theory are given in terms of a single non-perturbative parameter $\bar{\Lambda}$ and non-perturbative parameters of QCD, $\lambda_1$ and $\lambda_2$. In general, the mass of a hadron $H_Q$ containing a heavy quark Q obey an expansion of the form

$$m_X = m_Q + \bar{\Lambda} + \frac{\Delta m^2}{2m_Q} + O(\frac{1}{m_Q})$$

where X is the hadron, either in ground state (H) or an excited state (S), $m_Q$ is the mass of the heavy quark. X=H,S whereas $\Delta m^2 = -\lambda_1 + 2[J(J+1) - \frac{3}{2}]\lambda_2$. J is the total spin of meson. The two parameters $\lambda_1$ and $\lambda_2$ are non-perturbative parameters of QCD and can be estimated by fitting the theoretical and experimental data and their uncertainties [18,19]. A good estimation of these parameters may reduce theoretical errors and uncertainties up to significant level. Although there exists several predicted values in literature [20][21] for $\bar{\Lambda}$ and $\lambda_1$. In all cases, the values for $\lambda_1$ lie close to 1.0 GeV. The parameters can be fit by applying constraints through experimentally well defined masses and estimated parameter set can also be used to test the validity of other models and their predictions. The lowest and highest bounds on the parameters set can be found by using different values from the literature [22]. $\bar{\Lambda}$ and $\lambda_1$ can't be simply measured by mass measurements on dimensional grounds [23]. $\lambda_1$ is independent of $m_Q$ and $\lambda_2$ depends on $m_Q$ logarithmically. $\lambda_1$ and $\lambda_2$ are considered to possess same values for all states in a given spin-flavor multiplets and of the order of $\Lambda_{QCD}$ [23]. The term $\frac{\lambda_1}{m_Q}$ arises from kinetic energy of the heavy quark inside hadrons. The magnetic interaction $\lambda_2$ describes the interaction of the heavy quark spin with the gluon field and responsible for $B^* - B$ and $D^* - D$ splitting [23]. We here apply a suitable fitting procedure using Mathematica 7.0 to find the most suitable set of all the three parameters. The parameters are here allowed to vary with in their allowed values and then some of the sets that reproduce the masses with minimum error are chosen. One such set is shown as $\bar{\Lambda} = 0.6\, GeV$ which is close to global fitted value $0.57\pm0.07\, GeV$ given by [20] and $\lambda_1 = -0.18\pm0.06\, Gev^2$ for u and d light quarks. Assuming SU(3) breaking, $\bar{\Lambda}s = 0.7\, GeV$ and $\lambda_{1_s} = -0.18\pm0.06\, GeV^2$.

2. **Calculations and Result**

In the heavy quark limit, the heavy meson spectrum consists of degenerate heavy meson doublets where $J^P = 0^-, 1^-$ are the members of ground state doublet are and $J^P = 0^+, 1^+$ are the members of first excited state doublet. For the first excited and ground state doublet of bottom and charm mesons, the formula for the difference of spin-averaged masses of $J^P = 0^-, 1^-$ and $J^P = 0^+, 1^+$ states in the bottom sector is:

$$\overline{m}_S^{(Q)} - \overline{m}_H^{(Q)} = \overline{\Lambda}^S - \overline{\Lambda}^H - \frac{\lambda_1^S}{2m_Q} + \frac{\lambda_1^H}{2m_Q}$$

We use charm meson results to find the masses of higher bottom meson states, the hyperfine operators should be rescaled by $m_c/m_b$. This leads to the formula for splitting of the even and odd-parity states in the bottom sector.

$$\overline{m}_S^{(b)} - \overline{m}_H^{(b)} = \overline{m}_S^{(c)} - \overline{m}_H^{(c)} - (\lambda_1^S - \lambda_1^H)(\frac{1}{2m_c} - \frac{1}{2m_b}) \tag{1}$$

where $\quad \overline{m}_H^{(Q)} = (3m_{H^*}^{(Q)} + m_H^{(Q)})/4 \quad$ and $\quad \overline{m}_S^{(Q)} = (3m_{S^*}^{(Q)} + m_S^{(Q)})/4 \tag{2}$

Using $\lambda_1^H$ =-0.18±0.06GeV$^2$, $\lambda_1^{3/2} - \lambda_1^H$ =-0.23 GeV$^2$ [29], where $\lambda_1^{3/2}$ is the $\lambda_1$ matrix element for the j$^P$=3/2$^+$ doublet. We expect that the kinetic energy of the heavy quark in the j$^P$=1/2$^+$ states to be comparable to that of j$^P$=3/2$^+$ states. We take $\lambda_1^S - \lambda_1^H$ = -0.2±0.1GeV$^2$, and the mass of charm and bottom quarks, $m_c = 1.29^{+0.05}_{-0.11} GeV$, $m_b = 4.67 \pm 0.18\ GeV$ to find

$$\overline{m}_S^{(b)} - \overline{m}_H^{(b)} = \overline{m}_S^{(c)} - \overline{m}_H^{(c)} - 56.1 \pm 25 MeV.$$

$$\frac{m_{H^*}^{(b)} - m_H^{(b)}}{m_{H^*}^{(c)} - m_H^{(c)}} = \frac{m_{S^*}^{(b)} - m_S^{(b)}}{m_{S^*}^{(c)} - m_S^{(c)}} = \frac{m_c}{m_b} \tag{3}$$

Using Particle Data Group [28] we can calculate specifically their spin-averaged masses of the j$^P$=1/2$^-$ and 1/2$^+$ meson as

$$\overline{m}_{H_1}^c = (3m_{H_1^*}^{(c)} + m_{H_1}^{(c)})/4 = 0.03 \pm 0.2 MeV$$
$$\overline{m}_{H_3}^c = (3m_{H_3^*}^{(c)} + m_{H_3}^{(c)})/4 = 103.3 \pm 0.1 MeV$$
$$\overline{m}_{S_1}^c = (3m_{S_1^*}^{(c)} + m_{S_1}^{(c)})/4 = 447.95 \pm 23 MeV \tag{4}$$
$$\overline{m}_{S_3}^c = (3m_{S_3^*}^{(c)} + m_{S_3}^{(c)})/4 = 451 \pm 1.6 MeV$$

The hyperfine splitting of the B mesons is calculated as:

$$m_{H_1^*}^{(b)} - m_{H_1}^{(b)} = (\frac{m_c}{m_b})(m_{H_1^*}^{(c)} - m_{H_1}^{(c)}) = 39.03 \pm 0.12 MeV$$

$$m_{H_3^*}^{(b)} - m_{H_3}^{(b)} = (\frac{m_c}{m_b})(m_{H_3^*}^{(c)} - m_{H_3}^{(c)}) = 39.52 \pm 0.1 MeV$$

$$m_{S_1^*}^{(b)} - m_{S_1}^{(b)} = (\frac{m_c}{m_b})(m_{S_1^*}^{(c)} - m_{S_1}^{(c)}) = 39.72 \pm 0.5 MeV \tag{5}$$

$$m_{S_3^*}^{(b)} - m_{S_3}^{(b)} = (\frac{m_c}{m_b})(m_{S_3^*}^{(c)} - m_{S_3}^{(c)}) = 39.08 \pm 0.5 MeV$$

Using the values from the bottom non strange sector $m_{H_1}^{(b)} = 5279.1 \pm 0.4 \text{MeV}$ and $m_{H_1^*}^{(b)} = 5325.1 \pm 0.5 MeV$, $\bar{m}_{H_1}^{b}$ is found out to be $5313.62 \pm 0.03 \text{MeV}$. $m_{H_3}^{(b)} = 5366.3 \pm 0.6\ MeV$ is given in particle data group [28]. The value for $m_{H_3^*}^{(b)}$ will be:

$$m_{H_3^*}^{(b)} = (\frac{m_c}{m_b})(m_{H_3^*}^{(c)} - m_{H_3}^{(c)}) + m_{H_3}^{(b)} = 5404.82 \pm 0.7 \text{ MeV} \qquad (6)$$

From the relation (5) we get the spin-averaged masses of excited B-mesons
$$\bar{m}_{S_1}^{(b)} = (\bar{m}_{S_1}^{(c)} - \bar{m}_{H_1}^{(c)} - 56.1 \pm 25 MeV) + \bar{m}_{H_1}^{(b)} = 5705.44 \pm 48 MeV \qquad (7)$$

Similarly for the strange bottom and charm mesons, the strange bottom meson relation leads to
$$\bar{m}_{S_3}^{(b)} = (\bar{m}_{S_3}^{(c)} - \bar{m}_{H_3}^{(c)} - 56.1 \pm 25 MeV) + \bar{m}_{H_3}^{(b)} = 5691.8 \pm 27 MeV \qquad (8)$$

Equations (1) to (8) are solved to get the values for the masses of excited B mesons as shown in Table 1:

| Sr. No | State | Calculated Mass[MeV] | Experimental Value[28][MeV] | Potential Model[31][MeV] | Relativistic Model[30][MeV] |
|---|---|---|---|---|---|
| 1. | $m_{S_1}^{(b)}$ | $5691.6 \pm 345$ | $5366.7 \pm 0.24$ | 5697 | 5738 |
| 2. | $m_{S_1^*}^{(b)}$ | $5709.02 \pm 50$ | $5415.8 \pm 1.8$ | 5740 | 5757 |
| 3. | $m_{S_3}^{(b)}$ | $5662.48 \pm 27$ | ……... | 5716 | 5841 |
| 4. | $m_{S_3^*}^{(b)}$ | $5701.57 \pm 27$ | $5828.7 \pm 0.4$ | 5760 | 5859 |

Table 1: Calculated masses of excited B mesons compared with Experimental Values, Potential model and Relativistic model.

Comparison of our results with other models predicts the results to be matching well. In the charm and bottom systems, one knows experimentally [28]

$$\begin{aligned} m_{B^*} - m_B &\approx 46 MeV, \\ m_{D^*} - m_D &\approx 142 MeV, \\ m_{D_S^*} - m_{D_S} &\approx 142 MeV, \end{aligned} \qquad (9)$$

These mass splitting are in fact reasonably small. To be more specific, at order $1/m_Q$ one expects hyperfine corrections to resolve the degeneracy, for instance $m_{B^*} - m_B \propto 1/m_b$. This leads to the refined prediction $m_{B^*}^2 - m_B^2 \approx m_{D^*}^2 - m_D^2 \approx const$.

$$m_{B^*}^2 - m_B^2 \approx 0.8 GeV^2, \qquad m_{D^*}^2 - m_D^2 \approx 0.55 GeV^2 \qquad (10)$$

The spin symmetry also predicts that for strange mesons
$$m_{B_S^*}^2 - m_{B_S}^2 \approx m_{D_S^*}^2 - m_{D_S}^2 \approx const. \qquad (11)$$

But this constant could in principle be different from that for non strange mesons, since the flavor quantum numbers of the light degree of freedom are different in both cases. Experimentally, however,

$$m_{D_S^*}^2 - m_{D_S}^2 \approx m_{D^*}^2 - m_D^2$$

Indicating that to first approximation, hyperfine corrections are independent of the flavor of the "brown muck. One then expects the corresponding states in the bottom sector is

$$m_{B_2^*}^2 - m_{B_1}^2 \approx m_{D_2^*}^2 - m_{D_1}^2 \approx 0.17 GeV^2 \tag{12}$$

The fact that above mass splitting is smaller for the ground-state mesons is not unexpected. For instance, in the non-relativistic constituent quark model, the light antiquark in these excited mesons is in a p-wave state and its wave function at the location of the heavy quark vanishes. Hence, in this model hyperfine corrections are strongly suppressed. A typical prediction of the flavor symmetry is that the "excitation energies" for states with different quantum numbers of the light degrees of freedom are approximately the same in the charm and bottom systems. For instance, one expects

$$m_{B_S} - m_B \approx m_{D_S} - m_D \approx 100 MeV,$$
$$m_{B_1} - m_B \approx m_{D_1} - m_D \approx 557 MeV, \tag{13}$$
$$m_{B_2^*} - m_B \approx m_{D_2^*} - m_D \approx 593 MeV,$$

The first relation in Eq. [13] has been confirmed very nicely by the discovery of the $B_S$ meson by the ALEPH collaboration at Large Electron Positron Collider [32]. The observed mass, $m_{B_S} = 5.369 \pm 0.006 GeV$, corresponds to an excitation energy $m_{B_S} - m_B = 90 \pm 6 MeV$.

## 3. Conclusion

The spin-flavor symmetry in HQET leads to many interesting relations between the properties of hadrons containing a heavy quark. The most direct consequences concern the spectroscopy of such states. In the $m_Q \to \infty$ limit, the spin of the heavy quark and the total angular momentum j of the light degree of freedom are separately conserved by the strong interactions. Because of heavy quark symmetry, the dynamics is independent of the spin and mass of the heavy quark. Hadronic state can thus be classified by the quantum numbers (flavor, spin, parity) of the light degrees of freedom. The spin symmetry predicts that for fixed $j \neq 0$, there is a doublet of degenerate states with total spin $J \pm \frac{1}{2}$ and the flavor symmetry relates the properties of states with different heavy-quark flavor. This leads to the prediction that mass splitting among the various doublets are independent of heavy quark flavor. Our main purpose, in this paper is to find the masses of the non-strange excited states from the observed experimental values of all the ground states and excited strange mesons using the consequences of spin-flavor symmetry. The excited meson spectra calculated in the present paper is thus found to be matching well with other models [28-31]. Moreover, spin and flavor symmetry leads to various predictions for mass and hyperfine splitting. We also discuss briefly the predictions in equations (9-13) that can be made related to hyperfine splitting for strange as well as non-strange mesons. The importance of QCD parameters lies in the fact that it becomes comparatively easier to find using data of mass splitting and hyperfine splitting of heavy mesons. Since $\overline{\Lambda}$ has the same value for all particles in a spin-flavor multiplet, then $(\overline{\Lambda}^S - \overline{\Lambda}^H)$ can be taken to possess the same values for B and D mesons.

## 4. Acknowledgments